\documentclass[twocolumn,showpacs,preprintnumbers,amsmath,amssymb]{revtex4}

\usepackage{graphicx}% Include figure files
\usepackage{dcolumn}% Align table columns on decimal point
\usepackage{bm}% bold math

\begin{document}

\preprint{APS/123-QED}

\title{Cluster Structures in Oxygen isotopes}

\author{N. Furutachi} 
\author{S. Oryu}
\affiliation{Department of Physics, Faculty of Science and Technology, Tokyo University of Science, Noda, Chiba
278-8510, Japan}

\author{M. Kimura}
\affiliation{
Institute of Physics, University of Tsukuba, Tsukuba 305-8571, Japan
}%

\author{A. Dot\' e}
\affiliation{High Energy Accelerator Research Organization (KEK),
1-1 Ooho, Tsukuba, Ibaraki 305-0801, Japan.
}%

\author{Y. Kanada-En'yo}
\affiliation{Yukawa Institute for Theoretical Physics, Kyoto University, Kyoto 606-8502, Japan}

\date{\today}% It is always \today, today,
\begin{abstract}
Cluster structure of $^{16}$O,$^{18}$O and $^{20}$O is investigated 
by the antisymmettrized molecular dynamics (AMD) plus generator coordinate method (GCM). 
We have found the $K^{\pi}$=0$_2^+$ and 0$_1^-$ rotational bands of $^{18}$O that have the prominent
$^{14}$C+$\alpha$ cluster structure.
Clustering systematics becomes richer in $^{20}$O. We suggest the $K^{\pi}$=0$_2^+$ band that is the
mixture of the $^{12}$C+$\alpha$+4n and $^{14}$C+$^{6}$He cluster structures,
and the $K^{\pi}$=0$_1^-$ band that has the $^{14}$C+$^{6}$He cluster structure.
The $K^{\pi}$=0$_3^+$ and 0$_2^-$ bands that have the prominent $^{16}$C+$\alpha$ cluster structure
are also found.
\end{abstract}

\pacs{Valid PACS appear here}% PACS, the Physics and Astronomy
\maketitle

\section{Introduction}\label{int}
 
The investigation of the cluster structure of atomic nuclei has been mainly performed for light $N$=$Z$
nuclei, and much less is known for {\it N}$\neq${\it Z} and unstable nuclei. A famous example 
of the clustering in {\it N}$\neq${\it Z} nuclei is the 2$\alpha$ 
clustering of neutron-rich Be isotopes \cite{be1,be2,be3,amd-hf,be4,be5,be6,be7}.
The AMD studies \cite{be2,amd-hf,be4,be6} successfully described many properties of 
neutron-rich Be isotopes and showed that
most of them have the 2$\alpha$ cluster core. The degree of 2$\alpha$ clustering dynamically changes
depending on the motion of valence neutrons. This feature is well understood by the concept of 
the molecular-orbital \cite{be1,be3,be5}. 
The AMD study \cite{ne1} has predicted the existence of the cluster structures also in $^{22}$Ne, that is 
a system expected to have the $^{16}$O+$\alpha$
cluster core as an analogue of neutron-rich Be isotopes. In this study, it has been showed that the $^{16}$O+$\alpha$
core is formed and dissolved depending on the configurations of two valence neutrons.
Thus, the cluster structures of these nuclei suggests us 
richer variety of the cluster structure in {\it N}$\neq${\it Z} nuclei than {\it N}={\it Z} one.

To understand the clustering in {\it N}$\neq${\it Z} nuclei, a series of Oxygen isotopes is another good start point,
because the $^{12}$C+$\alpha$ cluster structure in 
$^{16}$O has been investigated in detail for a long time \cite{crrv}. Our interest in this study is 
to understand what will happen to the $^{12}$C+$\alpha$ cluster structure when we add neutrons to $^{16}$O.

Many of the low-lying excited states of $^{18}$O are understood within the shell model space
of two neutrons in $sd$-orbital above {\it N}={\it Z}=8 shell closure. However, it is also known that the states
with the core excitation ($4p2h$) coexist \cite{la,scc1} in the same energy region. 
They were associated to the $^{14}$C+$\alpha$ cluster structure, and many works were performed to investigate
the molecule band structure of $^{18}$O \cite{fas,mg1,mg3,mg2,scc1,scc2,ocm1,gcm2,gcm1}.
Consequently, the $K^{\pi}$=0$^+$ band that consists of the observed 
0$^+$ (3.63 MeV), 2$^+$ (5.26 MeV), 4$^+$ (7.11 MeV) and 6$^+$ (11.69 MeV) states has been 
established well as the $4p2h$ and $^{14}$C+$\alpha$ molecular band.
The cluster model calculations \cite{gcm2,gcm1} that showed the $^{14}$C+$\alpha$ cluster structure of the $K^{\pi}$=0$^+$ band
also predicted two molecular $K^{\pi}$=0$^-$ bands.
More recently, the $\alpha$-cluster structure was investigated by the 
elastic $\alpha$ scattering on $^{14}$C \cite{gol1} and $\alpha$ breakup reaction \cite{cur1,ash1,Yi1}.
In Ref. \cite{ash1}, it was proposed that the $K^{\pi}$=0$^-$ band built on the 1$^-$ state at 8.04 MeV \cite{cur1}, that
was in same energy region with the $K^{\pi}$=0$_2^-$ band predicted by the cluster models, was
the parity doublet partner of the $K^{\pi}$=0$^+$ $\alpha$-cluster band.
In the case of $^{20}$O, nothing is known about the clustering. In the investigation of the neutron transfer reaction
$^{18}$O$(t,p)$$^{20}$O compared with the shell model predictions, it has been suggested that the core excited states
($6p2h$) coexist with the normal states ($4p0h$) \cite{sm2}. We consider that these states (0$^+$(4.46 MeV), 2$^+$(5.30 MeV),
and 4$^+$(7.75 MeV)) could be associated with the cluster state as in the case of $^{18}$O. 

The purpose of this study is to investigate the cluster structure of $^{18}$O and $^{20}$O. 
We have applied the AMD+GCM (antisymmetrized molecular dynamics plus generator coordinate method)
framework. 
The AMD is a kind of {\it ab initio} theory in the sense that it can describe the cluster structure and
shell-like structure within the same framework without such assumptions as existence of clusters or an inert core.
Therefore this framework is useful to investigate the existence of the cluster structure in
$^{18}$O and $^{20}$O. We also calculate $^{16}$O to investigate how the $^{12}$C+$\alpha$ cluster states
are described in this framework.
We will suggest various kinds of cluster states in $^{18}$O and $^{20}$O. 
We find the $K^{\pi}$=0$^{\pm}$ bands that have prominent $^{14}$C+$\alpha$ cluster structure,
and some negative parity states that have non-negligible amount of $^{14}$C+$\alpha$ component.
In $^{20}$O, the motion of valence neutrons around the $^{12}$C+$\alpha$ cluster core enriches the variety of 
the clustering. Depending on the motion of valence neutrons, $^{12}$C+$\alpha$+4n, $^{14}$C+$^{6}$He and $^{16}$C+$\alpha$
cluster structures appear.

The contents of this article are as follows. In the next section, the AMD+GCM framework
is briefly outlined. In Sec. \ref{result}, the cluster structures of 
$^{16}$O, $^{18}$O and $^{20}$O are discussed. 
In the last section, we summarize this work.
 
\section{Theoretical framework}\label{fw}

\subsection{AMD wave function and calculational procedure}

In this subsection, the AMD+GCM framework is briefly outlined.
For more detail, readers are directed to Refs. \cite{amd2,amd3}. 
The AMD intrinsic wave function of $A$-nucleon system is described by a Slater determinant,
\begin{eqnarray}
\Phi_{{\rm int}} &=& \frac{1}{\sqrt{A!}} \det [ \varphi _1, \varphi _2, \cdots, \varphi _A ], \\
   \varphi _i (\mathbf{r}) &=& \phi _i (\mathbf{r}) \chi _i \tau _i  .   
\end{eqnarray}
Here, $\varphi _i $ is a single particle wave packet which is
composed of spatial part $\phi _i (\mathbf{r})$, spin part $\chi _{i}$,
and isospin part $ \tau _i $. The spatial part is described by a Gaussian,
\begin{eqnarray}
\phi_i (\mathbf{r}) &=& \left( \frac{2\nu}{\pi} \right )^{3/4}
\exp \left[ -\nu \left( \mathbf{r}-\frac{\mathbf{Z}_i}{\sqrt{\nu}} \right)^2
+\frac{\mathbf{Z}_i ^2}{2} \right].
\end{eqnarray}
where $\mathbf{Z}_i$ is complex three dimensional vector.
The width parameter $\nu$ is common for all nucleons and fixed to 0.17 fm$^{-2}$.
Spin part is parameterized by complex number parameter $\xi_i$,
\begin{eqnarray}
\chi _{i} &=& (\frac{1}{2}+\xi_i)\chi_{\uparrow} +(\frac{1}{2}-\xi_i)\chi_{\downarrow}.
\end{eqnarray}
The isospin part is fixed to up (proton) or down (neutron).
The $\mathbf{Z_i}$ and $\xi_i$ are the variational parameters and optimized by the
frictional cooling method.
The parity projected wave function which is generated from the $\Phi_{{\rm int}}$
is used as variational wave function,
\begin{eqnarray}
\Phi^{\pm}=\hat{P}^{\pm}\Phi_{{\rm int}}=\frac{(1\pm \hat{P}_x)}{2}\Phi_{{\rm int}},
\end{eqnarray}
where the $\hat{P}_x$ is the parity operator.

The Hamiltonian used in this study is given as,
\begin{eqnarray}
\hat{H}=\hat{T}+\hat{V}_{n}+\hat{V}_c-\hat{T}_g.
\end{eqnarray}
The $\hat{T}$ is the total kinetic energy and $\hat{T}_g$ is the energy of the center-of-mass motion, that is 
exactly treated in the AMD.
As the effective nuclear force $\hat{V}_{n}$, the Modified Volkov force (MV1) \cite{mv1} and spin-orbital
part of the G3RS \cite{g3rs} force are used. Details and the applied parameter set of these forces are
given in the next subsection.
Coulomb force $\hat{V}_c$ is approximated by a sum of
seven Gaussians.

The energy variation is performed under the constraint on the matter 
quadrupole deformation parameter $\beta$.
The constraint potential,
\begin{eqnarray}
V_{cnst}=\upsilon_{cnst}(\langle \beta \rangle - \beta_0)^2,
\end{eqnarray}
is added to the total energy of the system. Here, $\upsilon_{cnst}$ takes adequate positive value, 
and $\beta_0$ is a given number.
The definition of $\langle \beta \rangle$ is given in Ref. \cite{amd-hf}.

After the variation, the optimized wave function $\Phi^{\pm}(\beta)$ is projected to an eigenstate
of the total angular momentum $J$,
\begin{eqnarray}
\Phi_{MK}^{J\pm}(\beta)=\hat{P}_{MK}^{J}\Phi^{\pm}(\beta), 
\hspace{3mm} \hat{P}_{MK}^{J}=\int d\Omega D_{MK}^{J\ast}(\Omega)\hat{R}(\Omega).
\end{eqnarray}
The integrals over three eular angles are calculated numerically.

Finally, we superpose $\Phi_{MK}^{J\pm}(\beta)$ and diagonalize the Hamiltonian.
The wave function which describes a certain state is given as,
\begin{eqnarray}
\Psi_{n}^{J\pm}= \sum_i c_i^{n} \Phi_{MK_i}^{J\pm}(\beta_i),
\end{eqnarray}
where $c_i$ is determined by the Hill-Wheeler equation,
\begin{eqnarray}
\delta(\langle\Psi_n^{J\pm}|\hat{H}|\Psi_n^{J\pm}\rangle-\epsilon_n
\langle\Psi_n^{J\pm}|\Psi_n^{J\pm}\rangle)=0.
\end{eqnarray}

\subsection{Interactions}
We use the MV1 case3 force \cite{mv1} for central force, and the G3RS force \cite{g3rs} 
for spin-orbit force. 
The MV1 force consists of finite-range two-body and zero-range three-body terms,
\begin{eqnarray}
&&\hat{V}_{MV1}= \hat{V}^{(2)}+\hat{V}^{(3)}, \\
&&\hat{V}^{(2)}= \sum_{i<j}(1-m-m\hat{P}_{\sigma}\hat{P}_{\tau}) \nonumber \\
&&\times \{ \hat{V}_A \exp [ -\left( \hat{\mathbf{r}}_{ij}/r_A\right)^2 ] 
+ \hat{V}_R \exp [ -\left( \hat{\mathbf{r}}_{ij}/r_R\right)^2 ] \}, \\
&&\hat{\mathbf{r}}_{ij}=\hat{\mathbf{r}}_i-\hat{\mathbf{r}}_j, \\
&&\hat{V}^{(3)}=\sum_{i<j<k}\upsilon^{(3)}\delta(\hat{\mathbf{r}}_i-\hat{\mathbf{r}}_j)
\delta(\hat{\mathbf{r}}_i-\hat{\mathbf{r}}_k),
\end{eqnarray}
The spin-orbit part of the G3RS force is given as,
\begin{eqnarray}
\hat{V}_{LS}=\sum_{i<j} u \{ e^{ -\kappa_I \hat{\mathbf{r}}_{ij}^2 }
-e^{ -\kappa_{II} \hat{\mathbf{r}}_{ij}^2 } \}
\hat{P}(^3O)\hat{\mathbf{l}}_{ij}\cdot(\hat{\mathbf{s}}_i+\hat{\mathbf{s}}_j),
\end{eqnarray}
where $\hat{P}(^3O)$ is the projection operator onto the triplet odd state.
Adopted force parameters are summarized in Table. \ref{forcep}, and
the binding energies of $^{4}$He, $^{12,14}$C and $^{16,18,20}$O calculated using the 
present AMD+GCM framework are shown in Table. \ref{betb}.
For $^{12}$C, the excitation energies of the 2$_1^+$ and 4$_1^+$ states which are considered to be important
to describe the low-lying states of $^{16}$O are also shown. 

\begin{table}[h] 
\caption{The force parameters of the MV1 case3 force and spin-orbit part of the G3RS force. }\label{forcep}
\begin{ruledtabular}    
\begin{tabular}{cccccc}
 m & $V_A$ [MeV] & $V_R$ [MeV] & $r_A$ [fm] & $r_R$ [fm] & $\upsilon^{(3)}$ [MeV] \\
\hline
 0.61 & -83.34 & 104.86 & 1.60 & 0.82 & 4000 \\
\hline
\hline
\multicolumn{2}{c}{ u [MeV] } & \multicolumn{2}{c}{ $\kappa_I$ [fm$^{-2}$] } & \multicolumn{2}{c}{ $\kappa_{II}$ [fm$^{-2}$] } \\
\hline
 \multicolumn{2}{c}{3000} & \multicolumn{2}{c}{5.0} & \multicolumn{2}{c}{2.778}  
\end{tabular}
\end{ruledtabular}			
\end{table}

\begin{table}[h]  
\caption{The binding energies of $^{16,18,20}$O, $^4$He and $^{12}$C.
The excitation energies of 2$_1^+$ and 4$_1^+$ states of $^{12}$C are also shown.}\label{betb}
\begin{ruledtabular}
\begin{tabular}{ccccc}
                & \multicolumn{2}{c}{B.E [MeV] (Cal.)}     &  \multicolumn{2}{c}{B.E [MeV] (Exp.)} \\
      \hline
	$^4$He    &  \multicolumn{2}{c}{28.9}               &  \multicolumn{2}{c}{28.29}          \\
	$^{14}$C  &  \multicolumn{2}{c}{102.8}               &  \multicolumn{2}{c}{105.28}      \\
	$^{16}$O  &  \multicolumn{2}{c}{127.3}              &  \multicolumn{2}{c}{127.62}         \\
	$^{18}$O  &  \multicolumn{2}{c}{137.5}              &  \multicolumn{2}{c}{139.81}         \\
      $^{20}$O  &  \multicolumn{2}{c}{153.2}            &  \multicolumn{2}{c}{151.36}           \\
      \hline
      \hline
	    @    &         &  B.E [MeV](0$^+$)       & Ex.($2_1^+$) [MeV]  & Ex.($4_1^+$) [MeV]       \\
      \hline
	$^{12}$C &  Cal.    &  89.4                 & 5.5                      &  13.3             \\
	          &  Exp.   &  92.16                & 4.44                    &  14.1              \\
\end{tabular}		
\end{ruledtabular}			
\end{table}

\subsection{Analysis of the single-particle orbits}

We also investigate the single-particle structure of the obtained wave function $\Phi^{\pm}(\beta)$
by diagonalizing the single-particle Hamiltonian \cite{amd-hf}.
First, we transform the single-particle wave packets $\varphi_i$ into the orthonormal 
basis $\tilde{\varphi}_{\alpha}$,
\begin{eqnarray}
\tilde{\varphi}_{\alpha}=\frac{1}{\sqrt{\mu_{\alpha}}}\sum_{i=1}^A c_{i\alpha}\varphi_i.
\end{eqnarray}
Here $\mu_{\alpha}$ and $c_{i\alpha}$ are the set of eigenvalues and eigenvectors of the overlap matrix 
$B_{ij}\equiv\langle\varphi_i|\varphi_j\rangle$,
\begin{eqnarray}
\sum_{j=1}^{A}B_{ij}c_{j\alpha}=\mu_{\alpha}c_{i\alpha}.
\end{eqnarray} 
Using the $\tilde{\varphi}_{\alpha}$, we construct the single-particle 
Hamiltonian matrix,
\begin{eqnarray}
h_{\alpha\beta}&=&\langle\tilde{\varphi}_{\alpha}|\hat{t}|\tilde{\varphi}_{\beta}\rangle
+\sum_{i=1}^{A}\langle\tilde{\varphi}_{\alpha}\tilde{\varphi}_i|\hat{\upsilon}|\tilde{\varphi}_{\beta}\tilde{\varphi}_{i}
-\tilde{\varphi}_{i}\tilde{\varphi}_{\beta}\rangle \nonumber \\
&+&\frac{1}{2}\sum_{i,j=1}^{A}\langle\tilde\varphi_{\alpha}\tilde{\varphi}_{i}\tilde{\varphi}_{j}
|\hat{\upsilon}_3 | \tilde{\varphi}_{\beta}\tilde{\varphi}_{i}\tilde{\varphi}_{j}
+\tilde{\varphi}_{j}\tilde{\varphi}_{\beta}\tilde{\varphi}_{i} \nonumber \\
&+& \tilde{\varphi}_{i}\tilde{\varphi}_{j}\tilde{\varphi}_{\beta} 
-\tilde{\varphi}_{\beta}\tilde{\varphi}_{j}\tilde{\varphi}_{i} 
-\tilde{\varphi}_{i}\tilde{\varphi}_{\beta}\tilde{\varphi}_{j}
-\tilde{\varphi}_{j}\tilde{\varphi}_{i}\tilde{\varphi}_{\beta}\rangle,
\end{eqnarray}
Then we obtain the single-particle energy $\epsilon_p$ and single-particle wave function $\phi_s$
by the diagonalization of $h_{\alpha\beta}$,
\begin{eqnarray}
\sum_{\beta}h_{\alpha\beta}g_{\beta p}=\epsilon_p g_{\alpha p}, \\
\phi_s=\sum_{\alpha=1}^A g_{\alpha p}\tilde{\varphi}_{\alpha}.
\end{eqnarray}
In this study, we calculate the density distribution of the single-particle wave function $\phi_s$
to investigate the motion of valence neutrons.

\section{Results}\label{result}
\subsection{$^{16}$O}\label{seco16}

$^{16}{\rm O}$ is well known to have the prominent
$^{12}$C+$\alpha$ cluster structure in its excited
states \cite{crrv}. First, we investigate $^{16}{\rm O}$ and see how the
cluster structure is described in the present framework. 

\begin{figure}
	\includegraphics[scale=0.7]{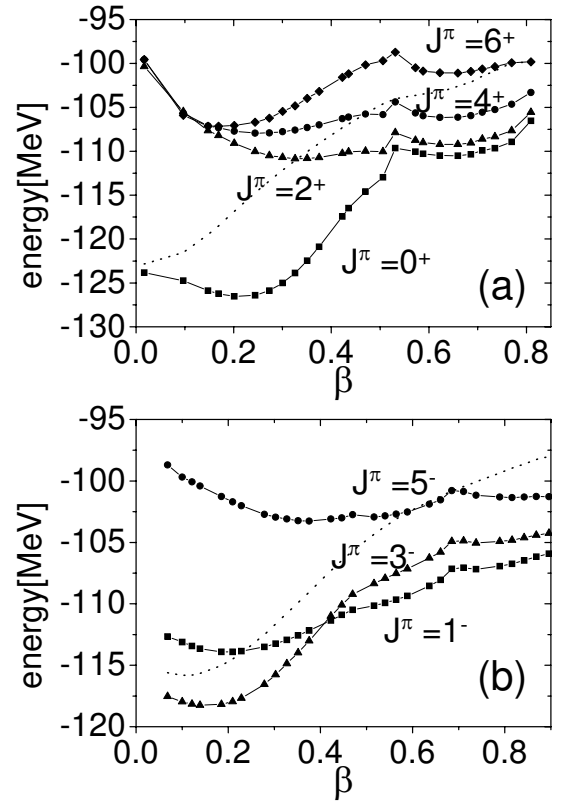}
	\caption{Energy curves of $^{16}$O as functions of matter quadrupole 
	         deformation parameter $\beta$ for the (a) positive- and (b) negative-parity states. 
	         Solid line represents the energy of each parity and
	         angular momentum ($K$=0) state, and dashed lines show the energy before the angular
	         momentum projection.}\label{o16es}
\end{figure}

\begin{figure}
      \includegraphics[scale=0.48]{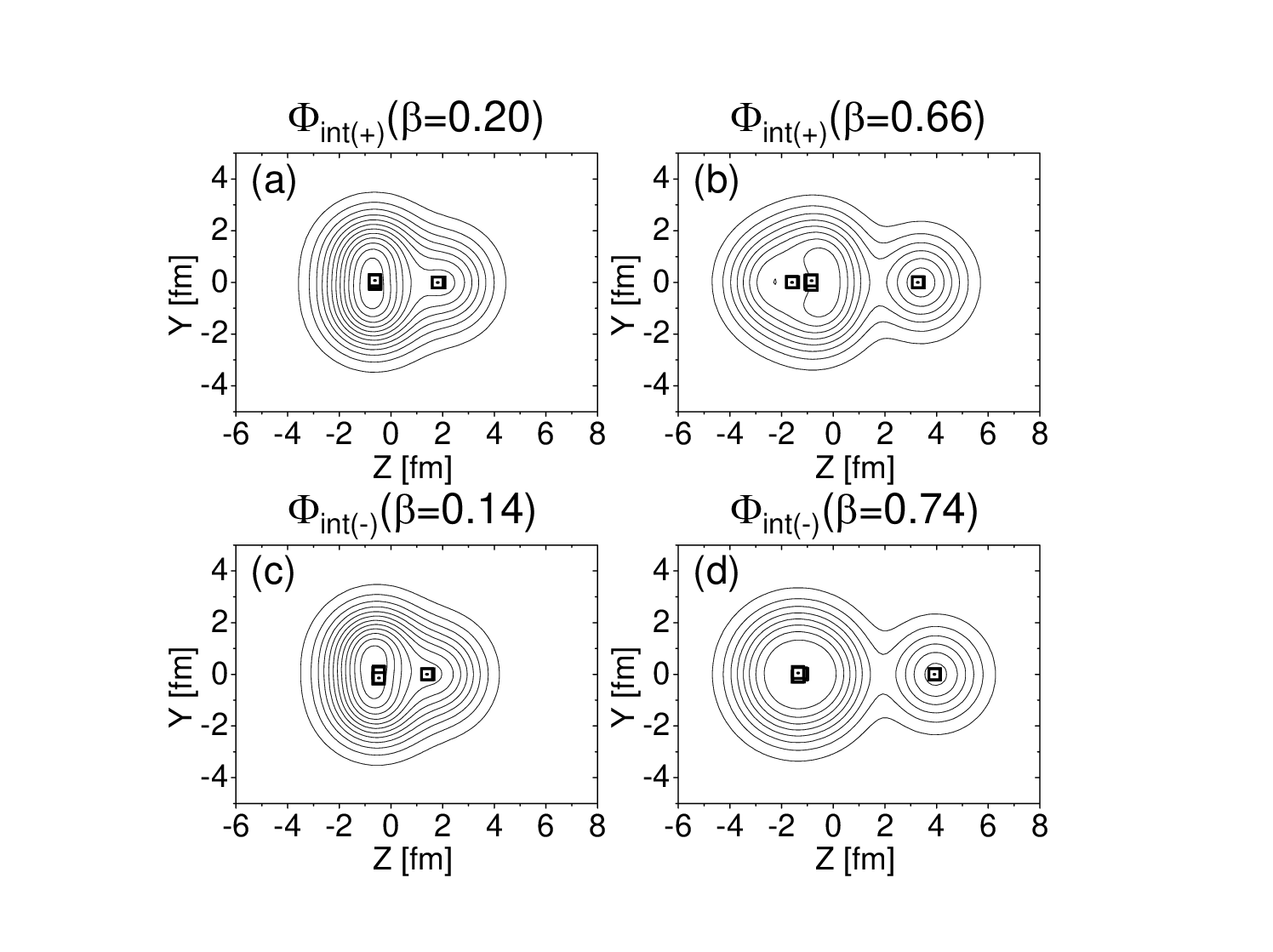}
	\caption{The matter density distributions 
	of the intrinsic wave functions of $^{16}$O.
	The centroids of the single-particle wave packets are plotted with
	white squares. The $\Phi_{int(+)}$ and $\Phi_{int(-)}$ denote the intrinsic wave function on the
	positive and negative parity curves, respectively.}\label{o16dens}
\end{figure}

The energy curves before and after the angular momentum
projection for the (a) positive- and (b) negative-parity states
are shown in FIG. \ref{o16es}. Before the angular momentum
projection, the positive-parity curve (dotted line in
FIG.\ref{o16es} (a)) has a energy minimum at the
spherical point. As the deformation becomes larger, the energy
rapidly increases. The angular momentum projection drastically changes the
energy curve. Here, we discuss each curve with respect to the angular momentum eigen states 
with $K$=0 for the sake of simplicity. 
The $0^+$ curve has a minimum at $\beta$=0.20 
and a shallow local minimum at $\beta$=0.66. 
The minimum state has the 0$\hbar \omega$ configuration and correspond to the ground state,
while the local minimum state has the 4$\hbar \omega$ configuration (proton 2$\hbar \omega$ and 
neutron 2$\hbar \omega$) and contributes to the 0$_2^+$ state.
Here the particle-hole configuration of each state is evaluated by 
the analysis of single particle orbits.
The density distributions of the intrinsic wave functions
at these minima are shown in FIG. \ref{o16dens} (a) and
(b). As clearly seen, the wave function at $\beta$=0.66 has the
prominent $^{12}$C+$\alpha$ clustering. In the case of the $2^+$,
$4^+$ and $6^+$ curves, they have two energy minima around
$\beta$=0.30 and 0.65. The energy curve of the
negative-parity state is also steep before the angular momentum
projection. After the angular momentum projection, the energy spectra
are different in the moderately deformed region ($\beta<0.5$)
and the largely deformed region ($\beta>0.5$). In the former region,
the lowest state is the $3^-$ state, and $1^-$ state is approximately 5 MeV above
the $3^-$ state. In the largely deformed region, the spectrum shows the
rotational nature. It is due to the structure change of the
intrinsic wave function. In the moderately deformed region, the wave
function has the 1$\hbar \omega$ configuration, while in the largely deformed
region, it has the $^{12}$C+$\alpha$ cluster structure as shown in FIG. \ref{o16dens} (c) and (d). 

\begin{figure}
	\includegraphics[scale=0.35]{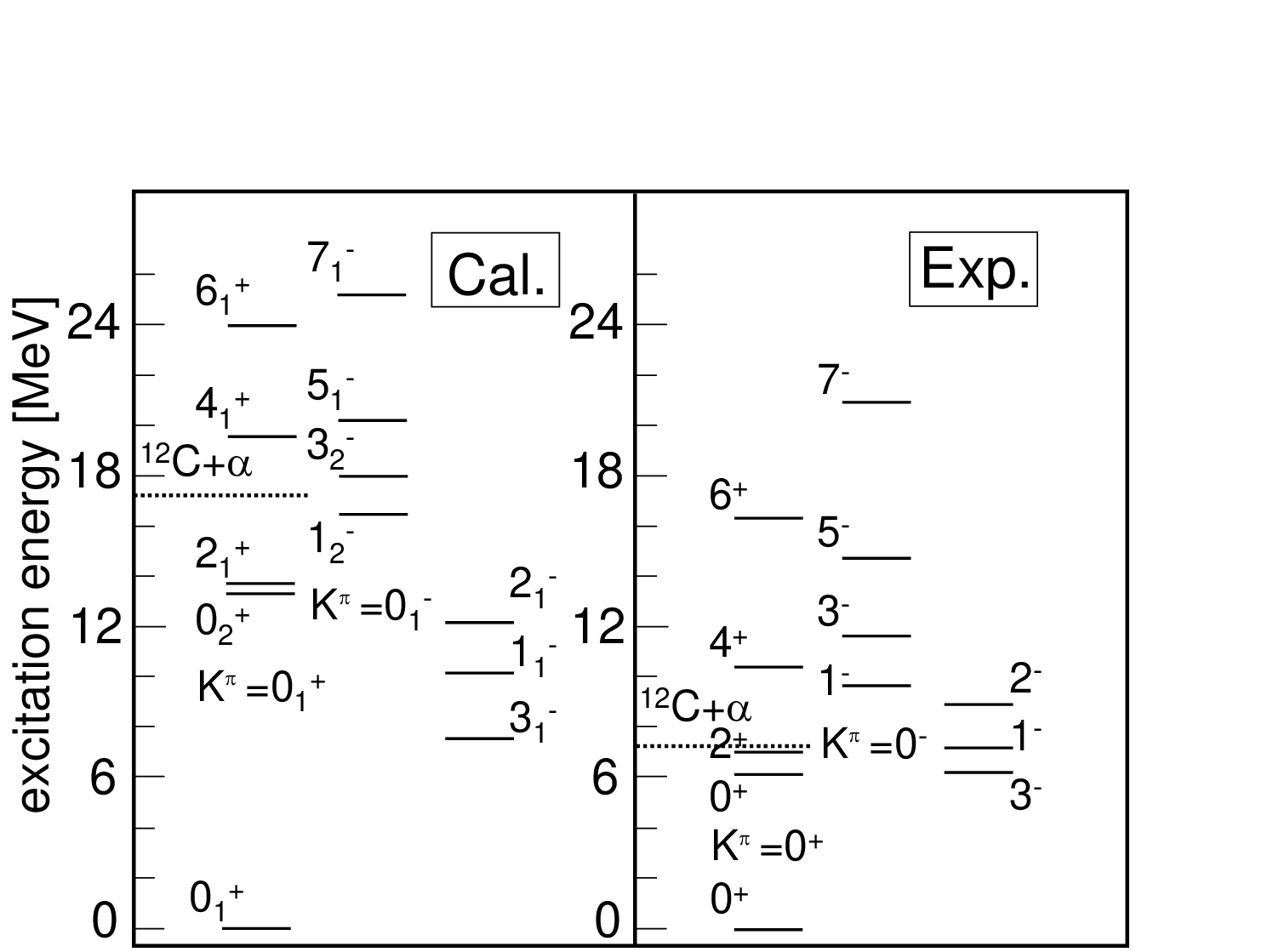}
	\caption{The excitation energies of the low-lying states of $^{16}$O. 
	Energies of the excited sates are shifted, for ease of comparison.
	Dotted lines show the theoretical and experimental threshold energies.}\label{o16level}
\end{figure}

After the angular momentum projection, we have performed the GCM calculation. 
The states with non-zero $K$ quantum number are also included into the GCM calculation.
FIG. \ref{o16level} shows the calculated and
observed level scheme. We have obtained the excited $K^{\pi}$=0$_1^+$ and 0$_1^-$
rotational bands together with the ground state and the low-lying $3^-_1$,
$1^-_1$ and $2^-_1$ states. The ground state dominantly consists of the
wave functions around the minimum at $\beta$=0.20. 
The excited $K^\pi$=0$_1^+$ and 0$_1^-$ rotational
bands mainly consist of the wave functions in the
largely deformed region ($\beta$=0.5-0.7 and $\beta$=0.6-0.8, respectively) that have the prominent
$^{12}$C+$\alpha$ cluster structure. The low-lying $3^-_1$, $1^-_1$
and $2^-_1$ states consist of the wave function that have the 1$\hbar \omega$
configuration. The excitation energy of the  $3^-_1$, $1^-_1$ and $2^-_1$ states
and the moment of inertia of the excited $K^\pi$=0$_1^+$ and 0$_1^-$ bands are
qualitatively reproduced. However, the calculated energy of the $K^{\pi}$=0$_1^+$ and 0$_1^-$ 
bands considerably overestimate the experimental value. 
We consider that one of the reason is that the internal wave function of $^{12}$C(g.s) cluster is not
correctly described in the present framework. In our wave function that
has the $^{12}$C+$\alpha$ cluster structure (FIG.\ref{o16dens} (b) and (d)),
the $^{12}{\rm C}$ is not in its ground state but has
some excited state component. This difficulty would be overcome if we performed the double angular momentum
projection (a simultaneous projection with respect to the $^{12}{\rm C}$ cluster and
to the total system), though it is rather time consuming. 
To compare with the experiment,
we utilize the $^{12}$C+$\alpha$ threshold energy evaluated without the projections.
Namely, we evaluate
it from the energies of $^{4}$He and $^{12}$C calculated without the parity, angular momentum projection and GCM, 
and call it theoretical threshold energy in the following.
The theoretical $^{12}$C+$\alpha$ threshold energy is 17.1 MeV (the energies of $^{12}$C and $^{4}$He 
are $-$83.5 and $-$26.7 MeV, respectively), and it is shown in FIG.\ref{o16level} by the dotted line.
When their energies are measured from the threshold, the $K^{\pi}$=0$_1^+$ and 0$_1^-$ bands show
better agreement with the experiment.
In the present results, the parity doublet bands constructed by the prominent $^{12}$C+$\alpha$ cluster structure
are consistent with the cluster model calculations \cite{crrv}. In addition, the single-particle excitations
in the low-lying negative parity states are also described.

\subsection{$^{18}$O}\label{seco18}

\begin{figure}
	\includegraphics[scale=0.7]{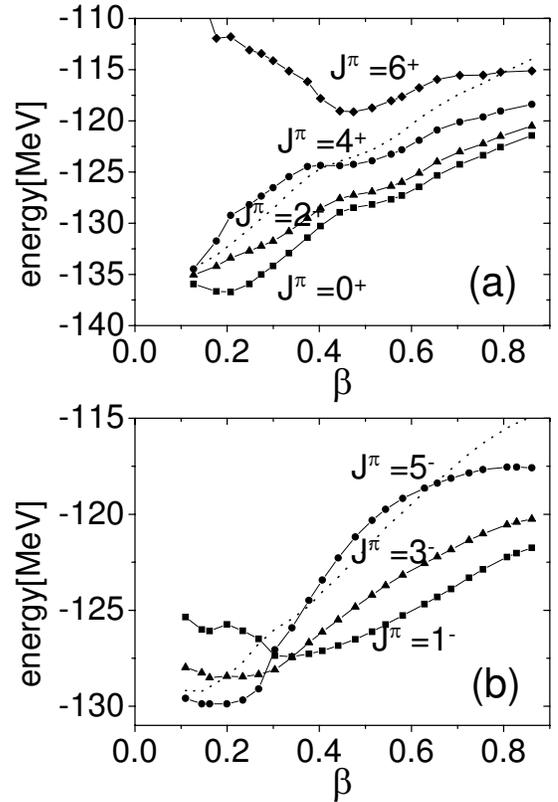}
	\caption{Energy curves of $^{18}$O as functions of matter quadrupole 
	         deformation parameter $\beta$ for the (a) positive- and (b) negative- parity states. 
	         Notations are same with FIG. \ref{o16es}.}\label{o18es}
\end{figure}

\begin{figure}
	\includegraphics[scale=0.47]{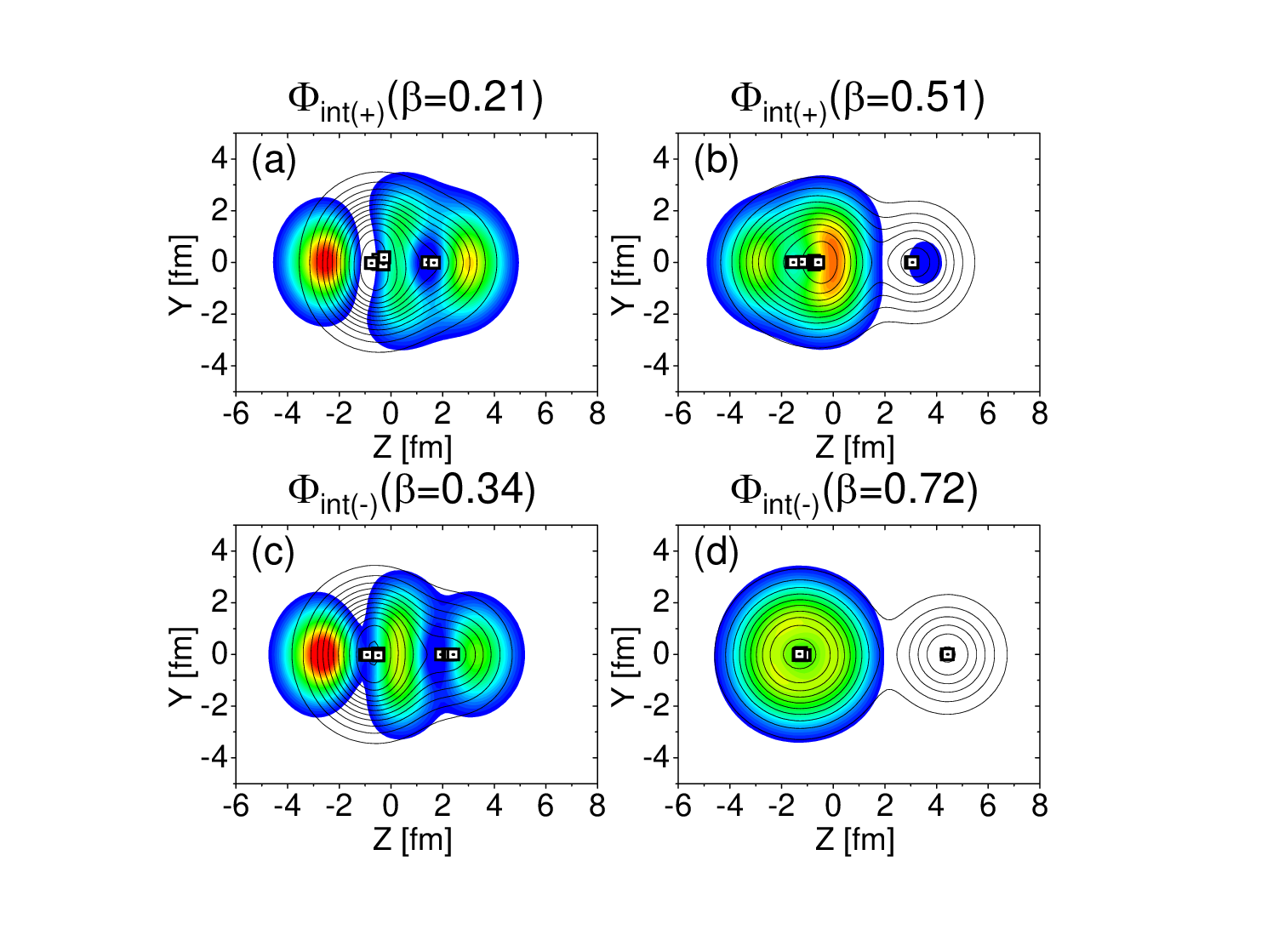}
	\caption{The density distributions of the core (black contour lines) 
	and the valence neutrons (color plots) of $^{18}$O. 
	The centers of the single-particle wave packets are shown by the white squares. 
      The intrinsic wave function of (a) and (c) gives the minimum energy for
	the 0$_1^+$ and 1$_1^-$ state, (b) and (d) becomes the dominant component of the 
	$K^{\pi}$=0$_2^+$ and $K^{\pi}$=0$_1^-$ rotational bands.}\label{o18dens}
\end{figure}

In this subsection, we investigate how the $\alpha$ cluster structure changes by adding two neutrons to $^{16}$O.
The same calculational procedure as the case of $^{16}$O is applied to $^{18}$O.
The obtained energy surface and the density distributions of the core and valence neutrons 
are shown in FIG. \ref{o18es} and \ref{o18dens}, respectively.
Here, we have defined the valence neutrons as two neutrons in the most weakly bound neutron orbitals,
and the core as the nucleons in the lowest 16 orbitals.
In the states shown in FIG. \ref{o18dens}, 
two valence neutrons occupy the orbitals that have the same spatial density distributions.
The 0$^+$ curve has the energy minimum at $\beta$=0.20, and 
the 2$^+$ and 4$^+$ curves have energy minima in the $\beta \sim 0$ region. 
These minimum states have the 0$\hbar \omega$ configuration, although the density distribution of the 
0$^+$ minimum state (FIG. \ref{o18dens} (a)) shows small deformation with the parity asymmetry.
Around $\beta$=0.45, the 6$^+$ curve has the energy minimum, 
and the 0$^{+}$, 2$^{+}$ and 4$^{+}$ curves have the shoulder. 
In this region, the wave functions are approximately correspond to the proton 2$\hbar \omega$ configuration.
The density distribution FIG. \ref{o18dens} (b) shows that the system is separated into two clusters.
There are 14 wave packets in the left side and 4 in the right side, that indicates the formation of the $^{14}$C+$\alpha$
cluster structure. Indeed, the density distribution of two valence neutrons  
shows that the valence neutrons stay only around the $^{12}$C cluster.
The formation of the $^{14}$C+$\alpha$ cluster structure leads to the rotational nature of the 0$^{+}$, 2$^{+}$,
4$^{+}$ and 6$^{+}$ energies.
The wave functions around $\beta$=0.51 become the dominant component of the 
$K^{\pi}$=0$_2^+$ rotational band after the GCM calculation.

In the case of the negative parity states (FIG. \ref{o18es}(b)), the 3$^{-}$ curve has the energy minimum at $\beta$=0.23, where
the wave function has the proton 1$\hbar \omega$ configuration. 
%It seems strange that the proton 1$\hbar \omega$ configuration makes lower energy for the 5$^-$ state. 
The 1$^-$ curve has the energy minimum at $\beta$=0.34. The density distribution of this state (FIG. \ref{o18dens} (c))
shows the slight development of the cluster structure. As deformation becomes larger, this cluster structure develops.
Figure \ref{o18dens} (d) shows the pronounced $^{14}$C+$\alpha$ cluster structure of the largely
deformed negative parity state.
Although the energy curves have no local minimum,
they become the dominant component of the $K^{\pi}$=0$_1^-$ rotational band after the GCM.
Again, the development of the $^{14}$C+$\alpha$ cluster structure is confirmed from 
the distributions of the wave packets and
the localization of the 
valence neutrons around $^{12}$C (FIG. \ref{o18dens} (d)).


\begin{figure}
	\includegraphics[scale=0.35]{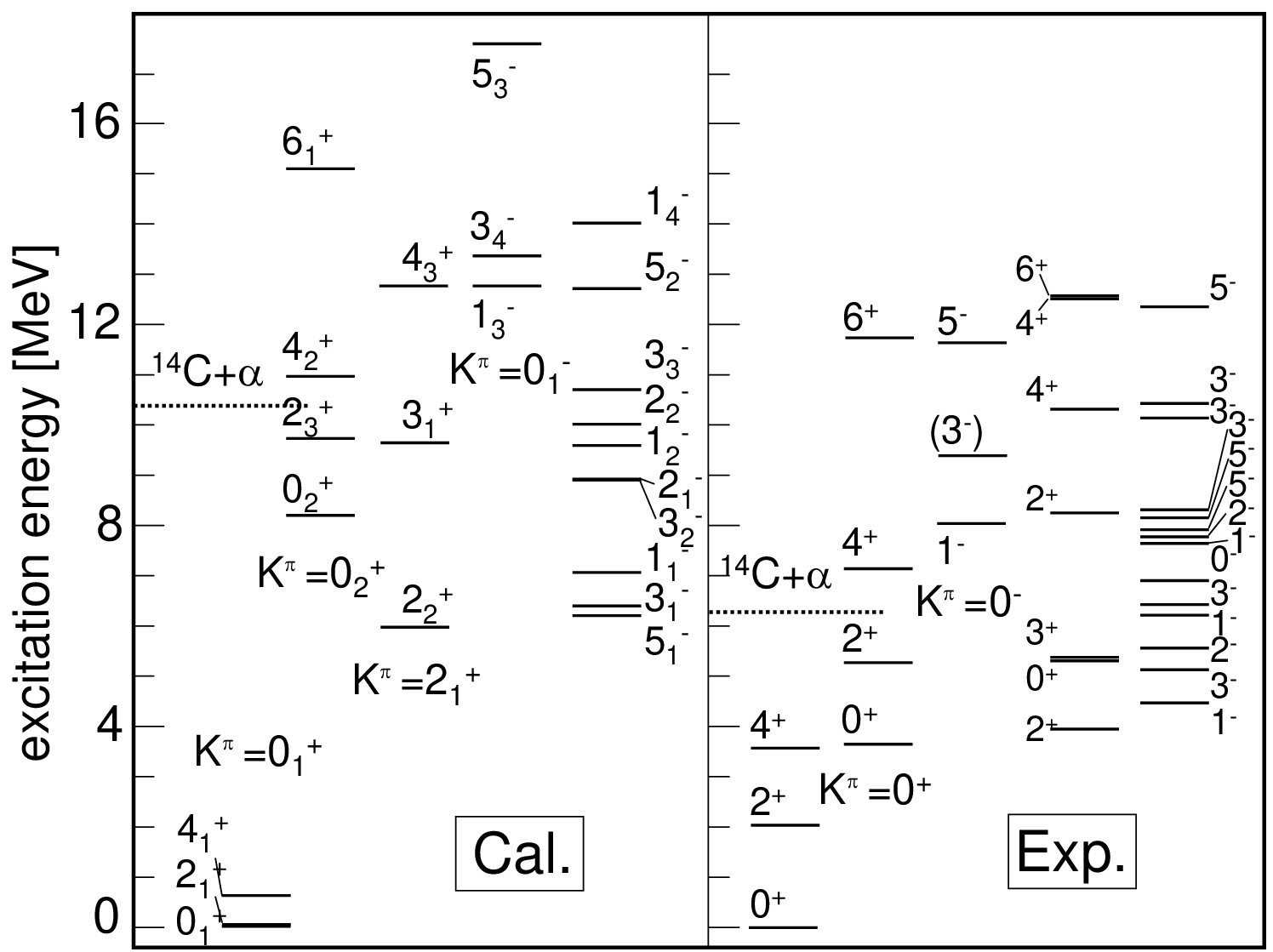}
	\caption{The low-lying level scheme of $^{18}$O.
	The experimental candidates of the $K^{\pi}$=0$_2^+$ and 0$_1^-$ bands are quoted from \cite{cu1,cur1}}\label{o18level}
\end{figure}

The level scheme obtained by the GCM calculation is shown in FIG. \ref{o18level} together with the experiment.
We have obtained the ground band and many excited states including the $K^{\pi}$=0$_2^+$ and 0$_2^-$ bands 
that have the $^{14}$C+$\alpha$ cluster structure.
The ground band (the 0$_1^+$, 2$_1^+$ and 4$_1^+$ states) dominantly consists of the wave functions
around $\beta$=0.20. Compared to the experimental spectrum, the level spacing in the ground band 
is considerably underestimated.
It is mainly due to the strong spin-orbit interaction used in this calculation. When
we weaken its strength as u=2000 MeV, the 2$_1^+$ and 4$_1^+$ states come to better
positions (2$_1^+$ state at 1.0 MeV and 4$_1^+$ state at 2.0 MeV).  
However, the use of the weaker spin-orbit interaction leads to the
overestimation of the $^{14}$C+$\alpha$ threshold energy.
Therefore we use the present strength parameter to discuss the $^{14}$C+$\alpha$ structure.

We have found that the 0$_2^+$, 2$_3^+$, 4$_2^+$ and 6$_1^+$ states dominantly consist of the wave functions 
around $\beta$=0.51, and hence classified them as the $K^{\pi}$=0$_2^+$ band.
As mentioned above, the intrinsic wave functions around $\beta$=0.51 have the $^{14}$C+$\alpha$ structure, therefore
this band is regarded to have the $^{14}$C+$\alpha$ cluster structure. 
The $^{14}$C+$\alpha$ threshold energy is overestimated as in the case of $^{16}$O. Therefore, 
to compare the obtained $K^{\pi}$=0$_2^+$ rotational band with the experiment, 
we measure its excitation energy from the threshold energy.
The theoretical $^{14}$C+$\alpha$ threshold energy calculated in the same way as the $^{12}$C+$\alpha$ 
threshold energy is 10.3 MeV 
(the energy of $^{14}$C is $-100.5$ MeV), while the experimental one is 6.24 MeV.
The energies of the 0$_2^+$, 2$_3^+$, 4$_2^+$ and 6$_1^+$ states measured from the $^{14}$C+$\alpha$ threshold energy are
$-2.1$, $-0.6$, 0.7 and 4.8 MeV, respectively. They
approximately agree with the experimental
0$^+$ (3.63 MeV), 2$^+$ (5.24 MeV), 4$^+$ (7.11 MeV) and 6$^+$ (11.69 MeV) states that are at 
$-2.61$, $-1.00$, 0.87 and 5.45 MeV measured from the threshold energy, respectively.
This assignment is consistent with that proposed in many theoretical works \cite{la,gcm2,gcm1,ocm1}
and the $\alpha$-transfer reaction \cite{cu1}.
We stress that the $^{14}$C+$\alpha$ cluster structure of the $K^{\pi}$=0$_2^+$ band has been found
without assuming any structure in our calculation.
We note that there is the mixing between the $^{14}$C+$\alpha$ cluster structure and the shell-like structure.
The $K^{\pi}$=2$_1^+$ band (the 2$_2^+$, 3$_1^+$ and 4$_3^+$ states) mainly consists of the wave functions around $\beta$=0.20, that have the 
0$\hbar \omega$ configuration. The natural parity states with this 
configuration are mixed with the $^{14}$C+$\alpha$ cluster wave functions ($K$=0). 
Therefore the 2$_2^+$ and 4$_3^+$ states also have considerable amount of the $^{14}$C+$\alpha$ component, although
the amount of their cluster component is smaller than that of the $K^{\pi}$=0$_2^+$ band.

The $K^{\pi}$=0$_1^-$ rotational band (1$_3^-$, 3$_2^-$ and 5$_2^-$)
dominantly consists of the wave functions around $\beta$=0.72, that have the
prominent $^{14}$C+$\alpha$ cluster structure. 
Therefore, the $K^{\pi}$=0$_1^-$ can be regarded as the parity doublet partner of the $K^{\pi}$=0$_2^+$ band.
The energies of the 1$_3^-$, 3$_2^-$ and 5$_2^-$ states measured from the theoretical $^{14}$C+$\alpha$ threshold energy are
2.5, 3.0, and 7.28 MeV, respectively. 
Experimentally, the assignment of the $K^{\pi}$=0$^-$ band which consists of the 1$^-$ (8.04 MeV), 3$^-$ (9.67 MeV), and 5$^-$ (11.62 MeV) sates was 
proposed by the $\alpha$ breakup reaction \cite{cur1}. The energies of these states measured from the 
$^{14}$C+$\alpha$ threshold energy are 1.80, 3.43, and 5.38 MeV, respectively.
We consider that they are candidates of the calculated $K^{\pi}$=0$_1^-$ band.
As in the case of the positive parity states, there is the mixing between the $^{14}$C+$\alpha$ structure and the proton
1$\hbar \omega$ configuration. This leads to the fragmentation of the $^{14}$C+$\alpha$ cluster structure
into many states. Especially the 1$_1^-$, 1$_2^-$, 3$_3^-$, 5$_2^-$ and 1$_4^-$ states have considerable $^{14}$C+$\alpha$ cluster state
component, though it is much smaller than that in the $K^{\pi}$=0$^-$ band members.

\subsection{$^{20}$O}\label{seco20}

\begin{figure}
	\includegraphics[scale=0.7]{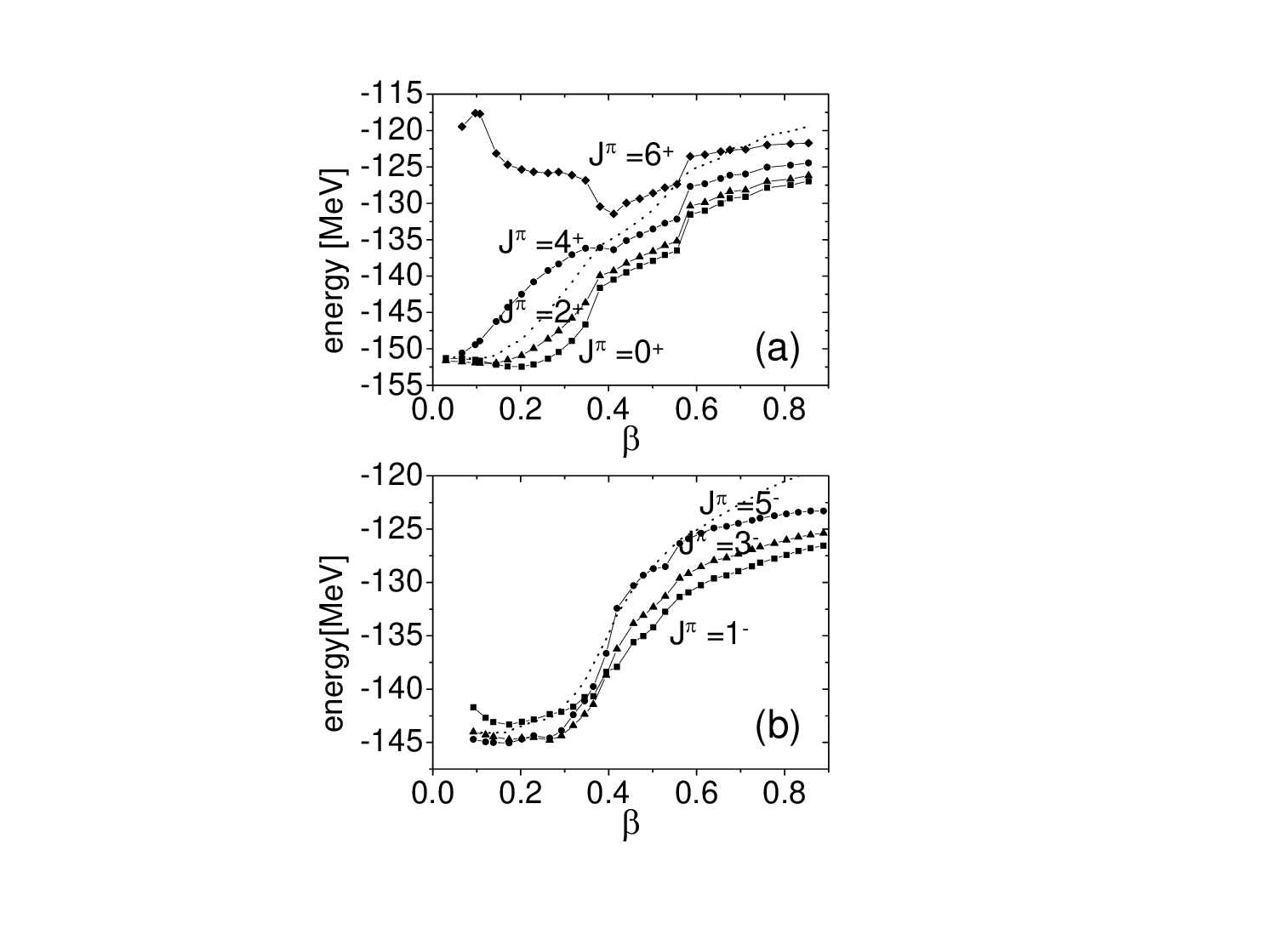}
	\caption{Energy curves of $^{20}$O as functions of matter quadrupole 
	         deformation parameter $\beta$ for the (a) positive- and (b) negative- parity states. 
	         Notations are same with FIG. \ref{o16es}}\label{o20es}
\end{figure}
\begin{figure*}
      \includegraphics[scale=0.6]{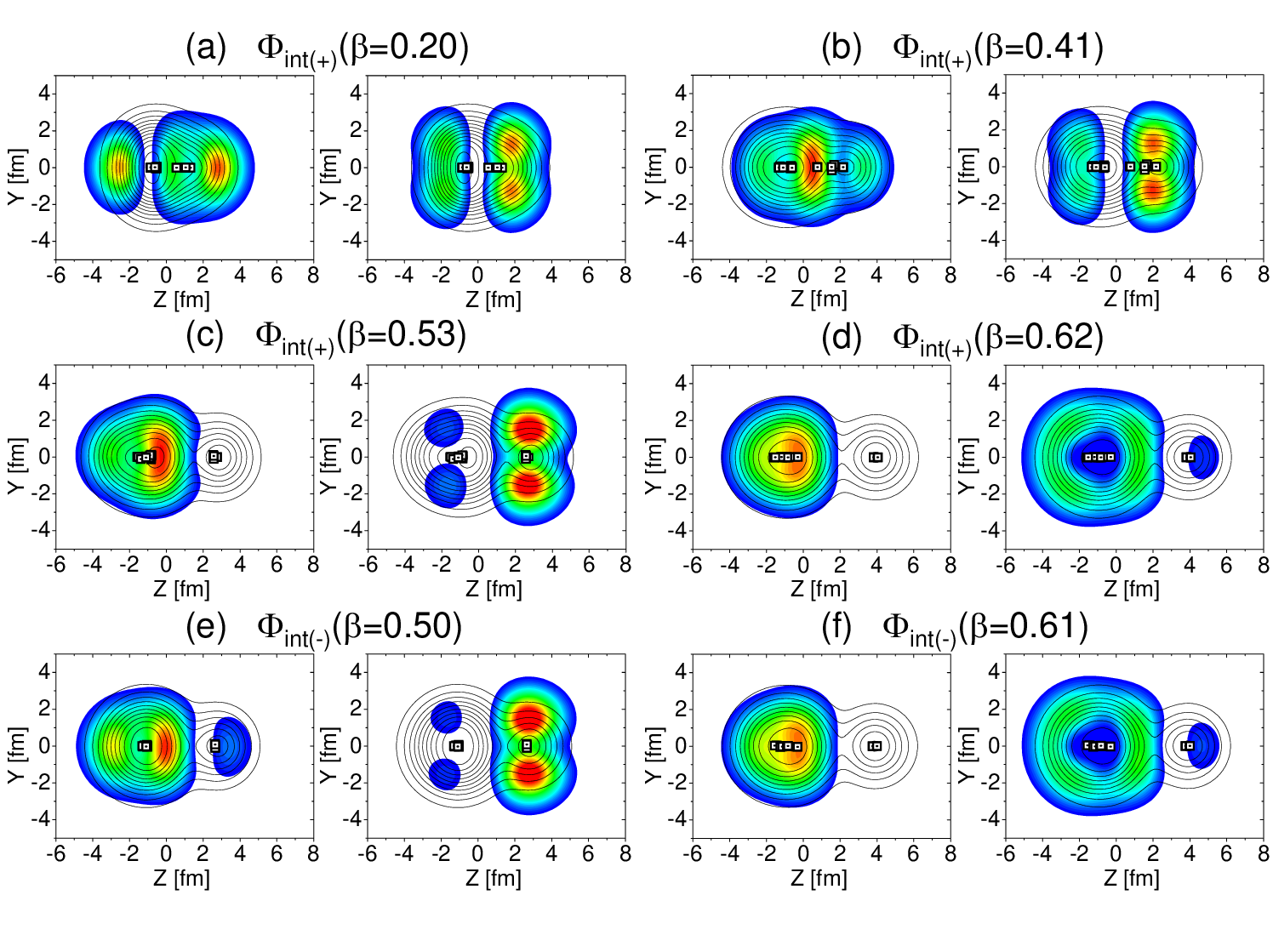}
      \caption{The density distributions of the core and the valence neutrons of $^{20}$O.
               Two valence neutron orbitals are shown for each intrinsic state.}\label{o20dens}
\end{figure*}
\begin{figure*}
      \includegraphics[scale=0.45]{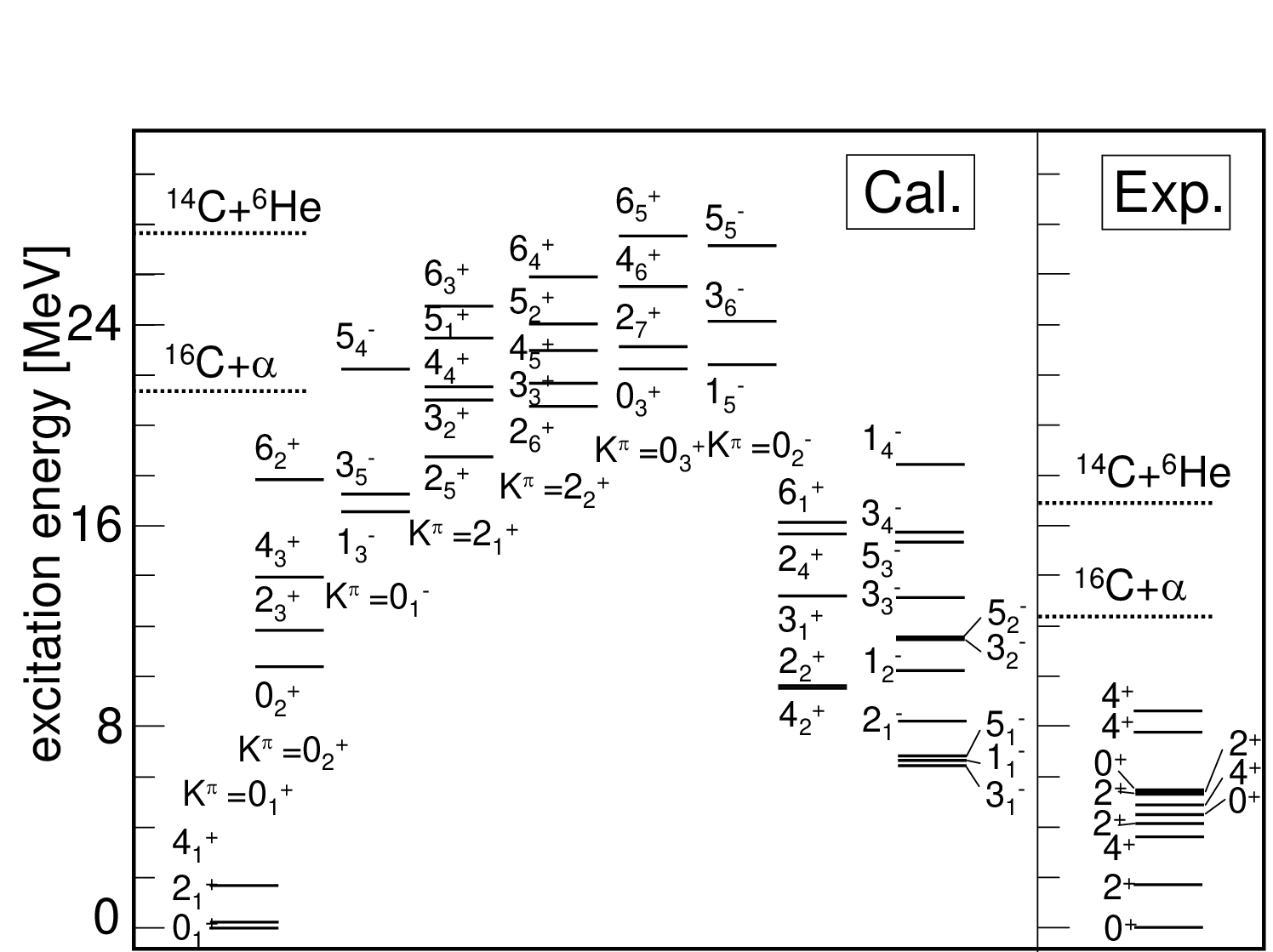}
	\caption{The level scheme of $^{20}$O.}\label{o20level}
\end{figure*}

By adding four neutrons to $^{16}$O, a variety of cluster states appears.
The obtained energy curves and the density distributions of the core and four valence neutrons are
shown in FIG. \ref{o20es} and \ref{o20dens}, respectively. 
We define the valence neutrons as the four neutrons in the most weakly bound neutron orbitals, and the core as
the nucleons in the lowest 16 orbitals.
In the states shown in FIG. \ref{o20dens}, there are always two orbitals that have different 
density distributions, and two valence neutrons occupy each orbital.
The 0$^+$ curve has the energy minimum at $\beta$=0.20 (FIG. \ref{o20dens} (a)), 
and the 2$^+$ and 4$^+$ curves have the 
energy minimum around $\beta$=0.10 and $\beta$=0, respectively. 
In this region, the wave functions have the 0$\hbar \omega$ configuration.
Around $\beta$=0.41, the structure changes from the 0$\hbar \omega$ to the proton 2$\hbar \omega$ configuration, 
and the 0$^+$ and 2$^+$ (4$^+$ and 6$^+$) curves have the shoulder (local minimum).
In this region, two different cluster structures appear.
Let us compare the wave functions 
at $\beta$=0.41 and $\beta$=0.53 (FIG. \ref{o20dens} (b) and (c)). 
Both of them have similar core density distributions which show
the development of the $^{12}$C+$\alpha$ cluster core. The difference between them is clearly seen in
the density distributions of the valence neutrons. 
The density distribution FIG. \ref{o20dens} (b) shows that four valence neutrons 
orbit around entire $^{12}$C+$\alpha$ core.
On the contrary, FIG. \ref{o20dens} (c) shows that two of four valence neutrons localize around
$^{12}$C cluster and the others localize around $\alpha$ cluster.
Therefore, we regard that the wave function at $\beta$=0.41 has the
$^{12}$C+$\alpha$+4n structure, and the wave function at $\beta$=0.53 has the $^{14}$C+$^{6}$He structure.
In the former, valence neutrons are moving in the mean field of the whole system of the $^{12}$C+$\alpha$ core,
while in the latter, the spatial correlations of two neutrons with the $^{12}$C and $\alpha$ core are enhanced.
From $\beta$=0.58, where the 0$^+$, 2$^+$, 4$^+$ and 6$^+$ curves have shoulder, another cluster structure appears.
The density distribution of the wave function at $\beta$=0.62 (FIG. \ref{o20dens} (d)) shows 
the formation of developed $^{16}$C+$\alpha$ structure, in which all valence neutrons orbit only around the $^{12}$C cluster.

Various structures also appear on the negative parity curve.
The 1$^-$, 3$^-$ and 5$^-$ curves have energy minimum around $\beta$=0.17, where the intrinsic wave functions
have the proton 1$\hbar \omega$ configuration. Around $\beta$=0.50, where the negative parity curves show the rotational nature,
the $^{14}$C+$^{6}$He structure appears. 
The density distributions of the core and valence neutrons in this state (FIG. \ref{o20dens} (e)) are quite similar to those of the $^{14}$C+$^{6}$He
structure that appears on the positive parity curve (FIG. \ref{o20dens} (c)).
In the largely deformed region, the $^{16}$C+$\alpha$ cluster structure appears around $\beta$=0.61 (FIG. \ref{o20dens} (f)), which is quite
similar to that found on the positive parity curve (FIG. \ref{o20dens} (d)).

The low-lying level scheme of $^{20}$O obtained by the GCM is shown in FIG. \ref{o20level} together with the experiment.
The ground band ($K^{\pi}$=0$_1^+$), many excited rotational bands ($K^{\pi}$=0$_2^+$, 0$_3^+$, 2$_1^+$, 2$_2^+$, 0$_1^-$ and 0$_2^-$) and other
excited states have been obtained.
The level spacing in the ground band (0$_1^+$, 2$_1^+$ and 4$_1^+$ states) is considerably underestimated compared to the 
experiment. It is due to the strong spin-orbit interaction, as in the case of $^{18}$O.
The $K^{\pi}$=0$_2^+$ band (0$_2^+$, 2$_3^+$, 4$_3^+$ and 6$_2^+$ states) dominantly consists of the wave functions
around $\beta$=0.41-0.58. In this region the $^{12}$C+$\alpha$+4n and $^{14}$C+$^{6}$He cluster structures appear,
as mentioned above. Therefore, the $K^{\pi}$=0$_2^+$ band is the mixture of these structures.
If we consider that the calculated 0$_2^+$ state correspond to the experimental 0$_2^+$ (4.46 MeV) state,
the experimental 2$^+$ (5.23 MeV) or 2$^+$ (5.30 MeV) state and 4$^+$ (7.75 MeV) 
states are the candidate of the calculated $K^{\pi}$=0$_2^+$ band from their energy positions.
In the experimental side \cite{sm2}, these 0$^+$ (4.46 MeV), 2$^+$ (5.30 MeV) and 4$^+$ (7.75 MeV) states 
have been assigned to the proton 2$\hbar \omega$ states predicted by the analysis with shell model calculations.
The $K^{\pi}$=2$_2^+$ band also consists of the wave functions around $\beta$=0.41.
The $K^{\pi}$=0$_1^-$ band (the 1$_3^-$, 3$_5^-$ and 5$_4^-$ states) dominantly consists of the wave functions 
around $\beta$=0.50, that have the $^{14}$C+$^{6}$He cluster structure.
Since the $^{14}$C+$^{6}$He cluster state component is contained also in the $K^{\pi}$=0$_2^+$ band,
the $K^{\pi}$=0$_2^+$ and 0$_1^-$ bands can be interpreted as the parity doublet bands, 
although the $K^{\pi}$=0$_2^+$ band has some $^{14}$C+$\alpha$+4n cluster state component.
In the negative parity states, there is the mixing between the proton 1$\hbar \omega$ and the $^{14}$C+$^{6}$He
state as in the case of $^{18}$O. This results in the fragmentation of the $^{14}$C+$^{6}$He cluster state component
into the 1$_4^-$, $3_4^-$ and 5$_3^-$ states.
Therefore these states have non-negligible amount of the $^{14}$C+$^{6}$He cluster state component.
The $K^{\pi}$=0$_3^+$ band dominantly consists of the wave functions around $\beta$=0.62, that have the 
$^{16}$C+$\alpha$ cluster structure. The $K^{\pi}$=0$_2^-$ band also consists of the wave functions 
that have the $^{16}$C+$\alpha$ cluster structure. 
As already mentioned, the $^{16}$C+$\alpha$ structures in the positive and
negative parity states around $\beta$=0.62 are quite similar to each other, therefore these bands are regarded to be the
parity doublet bands.
These bands start from just 
above the theoretical $^{16}$C+$\alpha$ threshold energy.
However, the experimental information on the levels around the $^{16}$C+$\alpha$ threshold (12.32 MeV)
is very little unfortunately. The $K^{\pi}$=2$_1^+$ band also consists of the wave functions around $\beta$=0.62.

Let us discuss characteristics of the cluster features of $^{20}$O in a series of O isotopes. 
As mentioned above,
we have found that the $\alpha$-cluster states follow the threshold rule systematically
in the series of $^{16}$O, $^{18}$O and $^{20}$O.
Namely, the $K^{\pi}$=0$_1^+$ band of $^{16}$O ($^{12}$C+$\alpha$ structure), $K^{\pi}$=0$_2^+$ band of $^{18}$O 
($^{14}$C+$\alpha$ structure) and $K^{\pi}$=0$_3^+$ band of $^{20}$O ($^{16}$C+$\alpha$ structure) appear
near the corresponding threshold energy.
On the contrary, the $K^{\pi}$=0$_2^+$ band of $^{20}$O appears at much smaller excitation energy
than the $^{14}$C+$^{6}$He 
threshold energy, although this band has large component of the $^{14}$C+$^{6}$He cluster structure.
We note that the $^{14}$C+$^{6}$He wave function (FIG. \ref{o20dens} (c)) has large overlap with the $^{12}$C+$\alpha$+4n
wave function (FIG. \ref{o20dens} (b)), and hence the system has molecular-orbital-like nature. 
Therefore, we consider that the valence neutrons play an important
role to lower the energy of the $K^{\pi}$=0$_2^+$ band.
The presence of the molecular-orbital-like band in $^{20}$O 
may be related to the weakly bound nature of $^{16}$C and $^6$He.
Since the last two neutrons in both nuclei are weakly bound, $^{16}$C+$\alpha$ and $^{14}$C+$^6$He do not appear
in small excitation energy. However, when two neutrons are covalently bound and shared by $^{14}$C and $\alpha$ 
clusters, it lowers the energy of the system.

\section{summary}
We have investigated the cluster structures of $^{16}$O, $^{18}$O and $^{20}$O, using the AMD+GCM 
framework. 
First, we have confirmed that the $K^{\pi}$=0$_1^+$ and 0$_1^-$ bands of $^{16}$O have the $^{12}$C+$\alpha$ 
cluster structure.
In $^{18}$O, The $K^{\pi}$=0$_2^+$ and 0$_1^-$ bands that have the $^{14}$C+$\alpha$ cluster structure are obtained.
They appear around $^{14}$C+$\alpha$ threshold energy and can be regarded as the parity doublet bands.
It is also noted that the $^{14}$C+$\alpha$ cluster structure is fragmented into many states. 
The valence neutrons give richer structure for $^{20}$O.
The analysis of the valence neutron orbitals revealed the presence of the cluster structures that have 
different motion of valence neutrons.
First is the $^{12}$C+$\alpha$+4n cluster structure, in which four valence neutrons orbit entire $^{12}$C+$\alpha$ core.
Second is the $^{14}$C+$^{6}$He cluster structure, in which the valence neutrons localize either of the $^{12}$C or 
$\alpha$ cluster. These structures are mixed and construct the $K^{\pi}$=0$_2^+$ band. 
This band has much smaller excitation energy than the $^{14}$C+$^{6}$He threshold energy.
The $^{14}$C+$^{6}$He cluster structure also constructs the $K^{\pi}$=0$_1^-$ band that can be regarded as 
the parity doublet partner of the $K^{\pi}$=0$_2^+$ band.
Third is the $^{16}$C+$\alpha$ cluster structure that constructs the $K^{\pi}$=0$_3^+$ and 0$_1^-$ 
parity doublet bands around the $^{16}$C+$\alpha$ threshold energy.
The appearance of the variety of cluster states in $^{20}$O may be related to the weak binding nature of subsystems, $^{16}$C and $^6$He.
They are not rigid cluster subunit, because the last two neutrons are weakly bound in both nuclei.
As a consequence, four valence neutrons orbit around $^{12}$C+$\alpha$ cluster core in different ways, and it leads to the
variety of cluster structures. It is an open problem whether the clustering phenomena appears in further neutron-rich
Oxygen isotopes.

\begin{acknowledgments}
We would like to thank Professor H. Horiuchi for valuable discussion. One of the authors (N.F.) also
thanks Professor K. Saito, Dr. T. Watanabe
and members of the nuclear theory group at Tokyo University of Science for various encouragement and discussions.
The numerical calculations were carried out on Altix3700 BX2 at YITP in Kyoto University.
\end{acknowledgments}

\end{document}